\documentclass[a4paper, 11pt]{article}
\usepackage[dvips,dvipdfm,dvipdfmx]{graphicx}
\usepackage{amsmath, amsthm}
\usepackage{amsmath}
\usepackage{amssymb}
\usepackage{mathrsfs}
\usepackage{bm}
\usepackage{colortbl}
\usepackage{multicol}
\usepackage{apalike}
\usepackage{tabularx} 

\newcolumntype{Y}{>{\centering\arraybackslash}X} 

\newcommand{\argmax}{\mathop{\rm arg~max}\limits}
\newcommand{\argmin}{\mathop{\rm arg~min}\limits}
\newcommand{\bs}{\boldsymbol}
\newcommand{\mb}{\mathbf}

\DeclareMathOperator{\sign}{sign}

\DeclareMathOperator{\df}{df}

\begin{document}

\begin{center}
\vspace{5.0cm}
{\huge Model-based clustering of multivariate binary data with dimension
 reduction}
\vspace{1.5cm}

{\Large
Michio Yamamoto \\
}
\vspace{0.5cm}
{
Kyoto University \\
}
\vspace{1.0cm}

{\Large
Kenichi Hayashi \\
}
\vspace{0.5cm}
{
Osaka University \\
}
\end{center}

\vspace{8.0cm}

Address correspondence to Michio Yamamoto, Department of Biomedical
Statistics and Bioinformatics, Kyoto University Graduate School of
Medicine, 54 Kawahara-cho, Shogoin, Sakyo-ku, Kyoto 606-8507, Japan.
Tel: +81-75-751-4745, Fax: +81-75-751-4732. Email:
michyama@kuhp.kyoto-u.ac.jp.

\newpage

\begin{abstract}
 Clustering methods with dimension reduction have been receiving
 considerable wide interest in statistics lately and a lot of methods to
 simultaneously perform clustering and dimension reduction have been
 proposed. This work presents a novel procedure for simultaneously
 determining the optimal cluster structure for multivariate binary data
 and the subspace to represent that cluster structure. The method is
 based on a finite mixture model of multivariate Bernoulli
 distributions, and each component is assumed to have a low-dimensional
 representation of the cluster structure. This method can be considered
 an extension of the traditional latent class analysis model. Sparsity
 is introduced to the loading values, which produces the low-dimensional
 subspace, for enhanced interpretability and more stable extraction of
 the subspace. An EM-based algorithm is developed to efficiently solve
 the proposed optimization problem. We demonstrate the effectiveness of
 the proposed method by applying it to a simulation study and real
 datasets.
\end{abstract}

\vspace{25pt}

\noindent Key words: Binary data; Clustering; Dimension reduction; EM
algorithm; Latent class analysis; Sparsity

\section{Introduction}

Binary data are commonly observed and analyzed in many application
fields: behavioral and social research, biosciences, document
classification, and inference on binary images. For example, Ekholm et
al. (2000) analyzed biomedical data including five unequally spaced
binary self-assessment measurements of arthritis and obesity data on the
presence or absence of obesity in five cohorts of children. Also, the
binarized data of the MovieLens 100K and the Netflix dataset, which are
popular datasets for collaborative filtering tasks, have been analyzed
by Kozma et al. (2009). One of the purposes of analyzing binary data, as
well as continuous data, is the partitioning of binary objects into
several unpredetermined homogeneous groups (clusters). For clustering of
multivariate data, it is quite important to know if some of the
variables do not contribute much to the structure of clusters because
the inclusion of redundant information can reduce the performance of the
cluster analysis (Milligan, 1996). Also, a lower-dimensional (say two or
three dimensional) representation of the cluster structure, based on the
most significant information, is very useful for evaluating and
interpreting the results of the cluster analysis.

Hence, what is needed is a procedure that constructs a low-dimensional
representation of the multivariate binary data, such that the cluster
structure in the data is maximally revealed.  For this purpose,
researchers often carry out a preliminary dimension reduction technique
(e.g., Collins et al., 2002; Schein et al., 2003; Lee et al.,
2010). Cluster analysis is then performed on the object scores on the
first few principal components. Although it is easy to implement, this
two-step sequential approach, also called the tandem approach, provides
no assurance that the components extracted in the first step are optimal
for the subsequent cluster analysis, because the two steps are
implemented separately by optimizing a different loss function (Arabie
and Hubert, 1994; DeSarbo et al., 1990; De Soete and Carroll, 1994;
Vichi and Kiers, 2001; Timmerman et al., 2010; Yamamoto and Hwang,
2014). For multivariate continuous data, instead of the two-step tandem
clustering procedure, several methods that simultaneously perform
cluster analysis and dimension reduction have been proposed (De Soete
and Carroll, 1994; Vichi and Kiers, 2001, Ghahramani and Hinton, 1997;
Yoshida et al., 2004).

On the other hand, for multivariate binary data, a few methods can
conduct the analysis for simultaneously obtaining a cluster structure
and a subspace for the cluster structure. Patrikainen and Mannila (2004)
have developed a subspace clustering method of binary data that can be
used in high-dimensional settings. Bouguila (2010) has developed a
clustering method for multivariate binary data with feature weighting
that allows variable selection taking variables with large
weights. Recently, Wu (2013) has proposed a penalized latent class model
for clustering extremely large-scale discrete data. This method can be
also considered a weighting method. Cagnone and Viroli (2012) have
proposed a factor mixture analysis model for multivariate binary data,
in which latent variables are distributed as a finite mixture of
multivariate Gaussian distributions.

In this paper, we focus on the common subspace clustering in which a
cluster structure is present in a low-dimensional space. As described
above, Patrikainen and Mannila's (2004) method allows for obtaining a
cluster structure and a subspace for the cluster structure
simultaneously. However, their method is rather cluster-specific
subspace clustering. In addition, in the past few decades, because of
technical advances in storing and processing data, we can obtain a large
dataset that includes a large number of variables. Thus, we need to take
into account such high-dimensional data. In a high-dimensional setting,
weighting methods for high-dimensional data, such as those of Bouguila
(2010) and Wu (2013), may be promising because variables that have lower
weights are suggested for exclusion from the model. However, their
methods do not provide explicit low-dimensional representation of the
data, which is useful for evaluating and interpreting the cluster
structure. Thus, in this paper, we propose a new method to
simultaneously find a cluster structure of multivariate binary data and
an optimal low-dimensional space for clustering. Furthermore, our
proposed method can deal with high-dimensional data.

The remainder of this paper is structured as follows. In Section
\ref{sec:proposed_method}, we propose a new method to cluster
multivariate binary data with dimension reduction. Section 3 describes
an algorithm for the proposed optimization problem. Section 4 is devoted
to studying the working of the clustering method using artificial and
real data examples. Finally, we sum up our findings and set out
directions for future expansion in Section 5.

\section{Proposed method}
\label{sec:proposed_method}

Let $\tilde{\bs{y}}=(\tilde{y}_{1},\dots,\tilde{y}_{D})'$ be a random
vector of $D$ binary variables. Suppose there are $K$ latent
(unobservable) classes in a population and let $\tilde{u}_{k}$,
$k=1,\dots,K$, be an allocation variable that takes ``1'' if an
observation belongs to class $k$, and ``0'' otherwise. We write
$\tilde{\bs{u}}=(\tilde{u}_{1},\dots,\tilde{u}_{K})'$. We assume that
the allocation variable follows a multinomial distribution, i.e., the
probability that $\tilde{\bs{u}}$ takes the value
$\bs{u}=(u_{1},\dots,u_{K})'$ is
\begin{equation*}
 f(\tilde{\bs{u}}=\bs{u})=\prod_{k=1}^{K}\xi_{k}^{u_{k}},
\end{equation*}
where
$\xi_{k}=\Pr(\tilde{u}_{1}=0,\dots,\tilde{u}_{k}=1,\dots,\tilde{u}_{K}=0)$.

Given that an observation is in the $k$th latent class, the probability
that the random vector $\tilde{\bs{y}}$ takes the value
$\bs{y}=(y_{1},\dots,y_{D})'$, where each $y_{d}$ takes $0$ or $1$, is
represented as $\Pr(\tilde{\bs{y}}=\bs{y}\mid \tilde{u}_{k}=1)$. The
unconditional probability of the response $\bs{y}$ when we do not know
the latent class of the observation is
\begin{equation}
 \Pr(\tilde{\bs{y}}=\bs{y})=\sum_{k=1}^{K}\xi_{k}\Pr(\tilde{\bs{y}}=\bs{y}\mid\tilde{u}_{k}=1).
	\label{eq:prob_y}
\end{equation}

Here, we need to specify how the probability
$\Pr(\tilde{\bs{y}}=\bs{y}\mid\tilde{u}_{k}=1)$ depends on
parameters. We postulate that, given the latent class to which an
observation belongs, the responses on the binary variables are
independent:
\begin{equation}
 \Pr(\tilde{\bs{y}}=\bs{y}\mid\tilde{u}_{k}=1)=
	\prod_{d=1}^{D}\Pr(\tilde{y}_{d}\mid\tilde{u}_{k}=1).
	\label{eq:prob_k}
\end{equation}
This assumption of {\it conditional independence} has been widely used
in latent class modeling in sociology (Collins and Lanza, 2010), and is
directly analogous to the assumption in the factor analysis model that
observed variables are conditionally independent given the factors
(Aitkin et al., 1981).

Finally, to specify the model completely, we need to specify a set of
parameters that define the conditional probability of $\tilde{\bm{y}}$,
with the value of $\tilde{\bm{u}}$ given. Suppose that
$\tilde{\bs{y}}_{1},\dots,\tilde{\bs{y}}_{N}$ are mutually independent
random variables that have the same distribution as $\tilde{\bs{y}}$,
and the entries of $\mb{Y}=(y_{nd})$ are those realizations. We assume
that, given the class $k$, $\tilde{y}_{d}$ follows the Bernoulli
distribution with success probability $\pi_{kd}$. For the traditional
latent class analysis model (Aitkin et al., 1981), we consider a
parameter vector $\bs{\theta}_{k}=(\theta_{k1},\dots,\theta_{kD})'$,
where $\theta_{kd}$ is the logit transformation of $\pi_{kd}$. We define
the inverse logit transformation
$\pi(\theta)=\{1+\exp(-\theta))\}^{-1}$. The success probabilities can
be represented using the canonical parameters $\theta_{kd}$ as
$\pi_{kd}=\pi(\theta_{kd})$. Let $\tilde{y}_{nd}$ be the $d$th element
of $\tilde{\bs{y}}_{n}$. The individual data-generating probability
given the class then becomes
\begin{align*}
 \Pr(\tilde{y}_{nd}=y_{nd}\mid\tilde{u}_{k}=1)
 &=\Pr(\tilde{y}_{nd}=y_{nd}\mid\tilde{u}_{k}=1,\theta_{kd})\\
 &=\pi(\theta_{kd})^{y_{nd}}\{1-\pi(\theta_{kd})\}^{1-y_{nd}}\\
 &=\pi(q_{nd}\theta_{kd}),
\end{align*}
with $q_{nd}=2y_{nd}-1$ since $\pi(-\theta)=1-\pi(\theta)$. Then, these
representations lead to the compact form of the log likelihood as
\begin{equation*}
 \sum_{n=1}^{N}\log\left(\sum_{k=1}^{K}\xi_{k}\prod_{d=1}^{D}\pi(q_{nd}\theta_{kd})\right).
\end{equation*}

We aim to obtain a low-dimensional representation of binary data in
which the true cluster structure exists. Thus, we assume that canonical
parameter $\theta_{kd}$ has a low-rank representation as follows:
\begin{equation}
 \theta_{kd}=\mu_{d}+\boldsymbol{f}_{k}'\boldsymbol{a}_{d},
	\label{eq:theta_assumption}
\end{equation}
where $\mu_{d}\in\mathbb{R}$, and for some positive integer $L$,
$\bs{f}_{k}\in\mathbb{R}^{L}$ and $\bs{a}_{d}\in\mathbb{R}^{L}$. Here,
$\mu_{d}$, $\bs{f}_{k}$, and $\bs{a}_{d}$ denote a centroid for the
$d$th variable, a component score of the $k$th cluster, and a loading
value for the $d$th variable, respectively. We write
$\bs{\xi}=(\xi_{1},\dots,\xi_{K})'$,
$\bs{\mu}=(\mu_{1},\dots,\mu_{D})'$,
$\mb{F}=(\bs{f}_{1},\dots,\bs{f}_{K})'$, and
$\mb{A}=(\bs{a}_{1},\dots,\bs{a}_{D})'$. To guarantee identifiability,
we require that $\mb{F}$ has orthonormal columns.  Then the log
likelihood can be written as
\begin{equation}
 \ell(\bs{\xi},\bs{\mu},\mb{F},\mb{A})=
	\sum_{n=1}^{N}\log\left(\sum_{k=1}^{K}\xi_{k}\prod_{d=1}^{D}
										 \pi(q_{nd}(\mu_{d}+\boldsymbol{f}_{k}'\boldsymbol{a}_{d}))\right).
	\label{eq:LL_theta_assumption}
\end{equation}

Here, to deal with the high-dimensional problem, we assume that most of
the elements of the true $\mb{A}$ are exactly zero. A sparse loading
matrix implies variable selection in cluster analysis. That is,
variables with non-zero loadings can be considered to contribute to a
cluster structure in a low-dimensional space, whereas variables with
zero loadings have no effect on the cluster structure. We propose to
perform variable selection using the penalized likelihood with
sparsity-inducing penalties. If $K=1$ and $\bs{f}_{k}$ is observable,
Eq. (\ref{eq:LL_theta_assumption}) is the log likelihood for $D$
logistic regression models. This connection with logistic regression
suggests the use of the $L_{1}$ penalty to obtain a sparse loading
matrix, as in the Lasso regression (Tibshirani, 1996). Specifically,
consider the penalty
\begin{equation*}
 P_{\lambda}(\mb{A})=\sum_{l=1}^{L}\lambda_{l}\|\check{\bs{a}}_{l}\|_{L_{1}}=
	\lambda_{1}\sum_{d=1}^{D}|a_{d1}|+\dots+\lambda_{L}\sum_{d=1}^{D}|a_{dL}|,
\end{equation*}
where $\check{\bs{a}}_{l}$ denotes the $l$th column of $\mb{A}$ and
$\lambda_{l}$ is a regularization parameter. The choice of values for
$\lambda_{l}$ will be discussed later. We obtain cluster components
$\bm{\xi}$, $\bm{\mu}$, and $\mb{F}$ and a sparse loading matrix
$\mb{A}$ by maximizing the following penalized log likelihood:
\begin{equation}
 S(\bs{\xi},\bs{\mu},\mb{F},\mb{A})=\ell(\bs{\xi},\bs{\mu},\mb{F},\mb{A})-N\cdot
	P_{\lambda}(\mb{A}).
	\label{eq:penLL}
\end{equation}

We call this procedure the clustering of binary data with reducing the
dimensionality (CLUSBIRD). We can interpret penalized maximization as
the device for generating a suitable optimization function, but not a
realistic representation of the actual data-generating process. Thus, in
this sense, the conditional independence given the latent class for
obtaining the likelihood in Eq. (\ref{eq:LL_theta_assumption}) is
assumed. A computational algorithm for solving the maximization problem
is presented in the next section.

\begin{figure*}[!tb]
 \begin{center}
\renewcommand{\arraystretch}{1}
 \vspace{0.5cm}
	\begin{tabular}{c}
	\includegraphics[width=10cm]{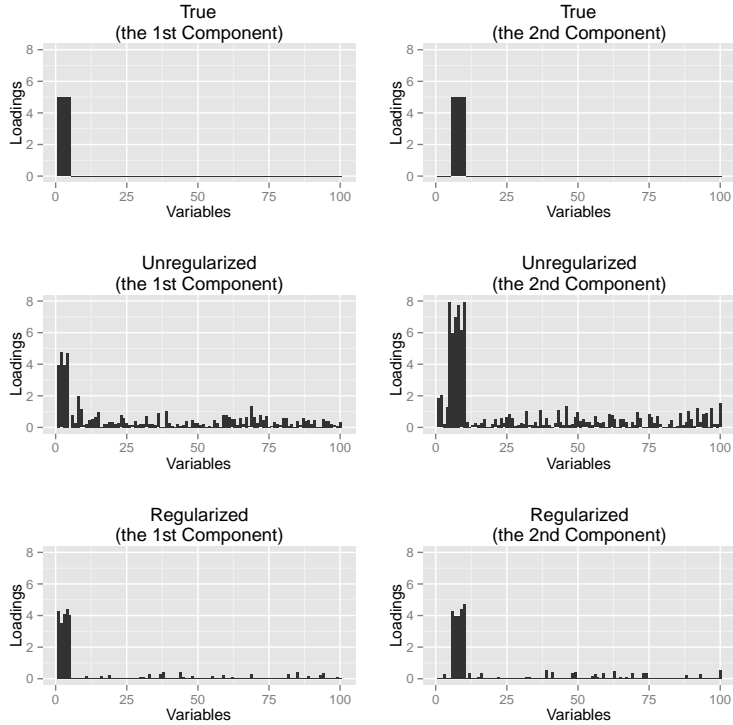}
	 \end{tabular}
 \caption{The results of analyzing an artificial dataset with $N=100$,
	$D=100$, $L=2$, and $K=4$; top, middle, and bottom panels show the
	true loadings, absolute values of loadings from the unregularized
	model, and absolute values of loadings from the regularized model,
	respectively; left and right panels show loadings for the first and
	second components, respectively; the penalty parameter was selected
	using the Bayesian information criterion} \label{fig:first_example}
\end{center}
\end{figure*}
The effectiveness of the introduction of sparsity is illustrated in
Figure \ref{fig:first_example} using a rank-two model (i.e., $L=2$). The
details of the setting will be presented in Section
\ref{sec:simulation}. While the regularized model can recover the
original loading vector efficiently under the sparsity assumption, the
unregularized model gives more noisy results. In the context of the
ordinary factor analysis model, a sparse structure for the loading
matrix provides an easy interpretation of the result, whereas it is
difficult to interpret the relation between variables and factors if the
loading matrix has no sparse structure. Browne (2001) provides an
excellent overview of the sparsity and rotation techniques which aim to
obtain a sparse structure. In addition, Hirose and Yamamoto (2014)
discuss the sparsity problem in the factor analysis model. Similar to
the ordinary factor analysis model, noisy loading values may lead to
difficulty in the interpretation of the result in our model. Thus, for
the proposed model, sparse loading values offer an advantage.

\section{Optimization Algorithm}

As is often the case, we apply the EM algorithm (Dempster et al., 1977)
to solve the maximization problem (\ref{eq:penLL}). Let
$\mb{U}=(u_{nk})$ be $N$ realizations of mutually independent random
variable $\tilde{\bs{u}}$. In addition, denote the conditional
probability (\ref{eq:prob_k}) by
$p_{k}(\bs{y}\mid\bs{\theta}_{k})$. Then, the complete-data likelihood
can be written as follows:
\begin{equation*}
 L^{C}(\mb{Y}, \mb{U}\mid\bs{\xi},\bs{\mu},\mb{F},\mb{A})
	=\prod_{n=1}^{N}\left\{\prod_{k=1}^{K}p_{k}(\bs{y}_{n}\mid\bs{\theta}_{k})^{u_{k}}
									 \prod_{k=1}^{K}\xi_{k}^{u_{k}}\right\}.
\end{equation*}
As described in the previous section, we aim to obtain the sparse
loading matrix $\mb{A}$; therefore, the penalty term for sparsity should
be introduced. Thus, the complete-data log-likelihood with the penalty
is
\begin{align}
 \ell^{C}(\mb{Y}, \mb{U}&\mid\bs{\xi},\bs{\mu},\mb{F},\mb{A})
 =\sum_{n=1}^{N}\sum_{k=1}^{K}u_{nk}\log p_{k}(\bs{y}_{n}\mid
	\bs{\theta}_{k})
 +\sum_{n=1}^{N}\sum_{k=1}^{K}u_{nk}\log\xi_{k}
	-N\cdot P_{\lambda}(\mb{A}).
	\label{eq:cll_penalty}
\end{align}

The EM algorithm consists of a step maximizing the conditional
expectation of the complete-data log-likelihood function
(\ref{eq:cll_penalty}) given the observable data $\mb{Y}$ and a set of
parameters,
$\{\bs{\xi}^{(t)},\bs{\mu}^{(t)},\mb{F}^{(t)},\mb{A}^{(t)}\}$. Here,
$\bs{\xi}^{(t)}$ denotes the value of $\bs{\xi}$ at the $t$th step in
the algorithm, and this notation is applied to other parameters. From
the above formulation, we can see that the penalized complete-data
log-likelihood (\ref{eq:cll_penalty}) is a linear function with respect
to values of $u_{nk}$. Thus, to obtain the conditional expected value of
$\ell^{C}$, we only have to replace $u_{nk}$ with its conditional
expectation,
\begin{align}
 u_{nk}^{*}&:=E\left[u_{nk}\mid\mb{Y};\bs{\xi}^{(t)},\bs{\mu}^{(t)},\mb{F}^{(t)},\mb{A}^{(t)}\right]\notag\\
	&=\frac{\xi_{k}^{(t)}p_{k}(\bs{y}_{n}\mid\bs{\theta_{k}}^{(t)})}
	{\sum_{k=1}^{K}\xi_{k}^{(t)}p_{k}(\bs{y}_{n}\mid\bs{\theta_{k}}^{(t)})},
	\label{eq:cond_expect}
\end{align}
where
$\bs{\theta}_{k}^{(t)}=(\theta_{k1}^{(t)},\dots,\theta_{kD}^{(t)})'$,
$k=1,\dots,K$, is obtained through Eq. (\ref{eq:theta_assumption}) using
$\{\bs{\xi}^{(t)},\bs{\mu}^{(t)},\mb{F}^{(t)},\mb{A}^{(t)}\}$.  Thus,
the conditional expectation of the complete-data log-likelihood is as
follows:
\begin{align*}
 Q(\bs{\xi},\bs{\mu},\mb{F},\mb{A}&\mid\bs{\xi}^{(t)},\bs{\mu}^{(t)},\mb{F}^{(t)},\mb{A}^{(t)})\\
 &=E\left[\ell^{C}\mid\mb{Y};\bs{\xi}^{(t)},\bs{\mu}^{(t)},\mb{F}^{(t)},\mb{A}^{(t)}\right]\\
 &=\sum_{n=1}^{N}\sum_{k=1}^{K}u_{nk}^{*}\log
 p_{k}(\bs{y}_{n}\mid\bs{\theta}_{k})
 +\sum_{n=1}^{N}\sum_{k=1}^{K}u_{nk}^{*}\log\xi_{k}
 -N\cdot P_{\lambda}(\mb{A}).
\end{align*}

In the M-step of the EM algorithm, we consider the following maximization
problem
\begin{equation}
 (\hat{\bs{\xi}},\hat{\bs{\mu}},\hat{\mb{F}},\hat{\mb{A}})=
	\argmax_{\bs{\xi},\bs{\mu},\mb{F},\mb{A}}Q(\bs{\xi},\bs{\mu},\mb{F},\mb{A}\mid\bs{\xi}^{(t)},\bs{\mu}^{(t)},\mb{F}^{(t)},\mb{A}^{(t)}).
	\label{eq:max_Q}
\end{equation}

Same as the usual mixture models, the estimate of $\bs{\xi}$ can be
obtained by
\begin{equation}
 \hat{\xi}_{k}=N^{-1}\sum_{n=1}^{N}u_{nk}^{*},\ \ \text{for $k=1,\dots,K-1$},
	\label{eq:update_xi}
\end{equation}
and $\hat{\xi}_{K}=1-\sum_{k=1}^{K-1}\hat{\xi}_{k}$.

Given the estimate of $\bs{\xi}$, the maximization problem in
(\ref{eq:max_Q}) with respect to $\bs{\mu}$, $\mb{F}$, and $\mb{A}$ is
equivalent to the minimization of the following function:
\begin{equation}
 g(\bs{\mu},\mb{F},\mb{A})=-\sum_{n=1}^{N}\sum_{k=1}^{K}u_{nk}^{*}
	\log p_{k}(\bs{y}_{n}\mid\bs{\theta}_{k})+N\cdot P_{\lambda}(\mb{A}).
	\label{eq:min_g}
\end{equation}
Here, the function $g$ in (\ref{eq:min_g}) is non-quadratic. Then,
instead of directly dealing with the non-quadratic function $g$, we
minimize a surrogate function, called the majorizing function (Hunter
and Lange, 2004), to solve the minimization problem of a quadratic
function. In the majorization algorithm, a suitably defined quadratic
upper bound of (\ref{eq:min_g}) is minimized, which provides optimal
values for the actual function $m$. A function $h(x\mid y)$ is said to
majorize a function $m(x)$ at $y$ if
\begin{equation*}
 h(x\mid y)\ge m(x)\ \ \ \text{for all $x$}\ \ \ \text{and}\ \ \ h(y\mid
	y)=m(y).
\end{equation*}
In the geometrical view, the function surface $h(x\mid y)$ lies above
the function $m(x)$ and is tangent to it at the point $y$; therefore
$h(x\mid y)$ becomes an upper bound of $m(x)$. To minimize $m(x)$, the
majorization algorithm decreases the objective function $m(x)$ in each
step and is guaranteed to converge to a local minimum of $m(x)$. When
applying the majorization algorithm, the majorizing function $h(x\mid
y)$ is chosen so that it is easier to minimize than the original
objective function $m(x)$. The study by Hunter and Lange (2004) can be
referred for an introductory description of the majorization algorithm.

To find a suitable majorizing function of (\ref{eq:min_g}), we consider
the first term of (\ref{eq:min_g}). Note that, for a given point $y$,
\begin{equation}
 -\log\pi(x)\le-\log\pi(y)-\{1-\pi(y)\}(x-y)+\frac{1}{8}(x-y)^{2},
	\label{eq:inequality1}
\end{equation}
and the equality holds when $x=y$ (Jaakkola and Jordan, 2000; De Leeuw,
2006). This equation provides quadratic upper bounds for the first term
of (\ref{eq:min_g}) at the tangent point $y$.  Thus we can apply the
majorization algorithm for our problem.

We now present details of the majorization algorithm via the upper bound
of $-\log\pi(x)$ in (\ref{eq:inequality1}). By completing the square,
Eq. (\ref{eq:inequality1}) can be rewritten as
\begin{equation}
 -\log\pi(x)\le-\log\pi(y)+\frac{1}{8}\left[x-y-4\left\{1-\pi(y)\right\}\right]^{2}.
	\label{eq:upper_bound2}
\end{equation}
Substituting $x$ and $y$ with $q_{nd}\theta_{kd}$ and
$q_{nd}\theta_{kd}^{(t)}$, respectively in (\ref{eq:upper_bound2}) and
using $q_{nd}=\pm1$, we obtain
\begin{equation}
 -\log\pi(q_{nd}\theta_{kd})\le-\log\pi(q_{nd}\theta_{kd}^{(t)})+
	\frac{1}{8}(\theta_{kd}-z_{nkd}^{(t)})^{2},
\end{equation}
where
\begin{equation*}
 z_{nkd}^{(t)}=\theta_{kd}^{(t)}+4q_{nd}\left\{1-\pi(q_{nd}\theta_{kd}^{(t)})\right\}.
\end{equation*}
Thus, we obtain the following quadratic upper bound of the first term of
(\ref{eq:min_g}):
 \begin{equation}
	\frac{1}{8}\sum_{n=1}^{N}\sum_{k=1}^{K}u_{nk}^{*}\sum_{d=1}^{D}(\theta_{kd}-z_{nkd}^{(t)})^{2}.
	 \label{eq:upper_bound3}
 \end{equation}
Eq. (\ref{eq:upper_bound3}) then yields the following upper bound (up to
a constant) of the criterion function $g(\bs{\mu},\mb{F},\mb{A})$
defined in (\ref{eq:min_g}):
 \begin{align}
	h(\bs{\mu},&\mb{F},\mb{A}\mid\bs{\mu}^{(t)},\mb{F}^{(t)},\mb{A}^{(t)})\notag\\
	&=\frac{1}{8}\sum_{n=1}^{N}\sum_{k=1}^{K}u_{nk}^{*}
	 \|\bs{z}_{nk}^{(t)}-(\bs{\mu}+\mb{A}\bs{f}_{k})\|^{2}
	 +N\cdot P_{\lambda}(\mb{A}),
	\label{eq:func_h}
 \end{align}
 where $\bs{z}_{nk}^{(t)}=(z_{nk1}^{(t)},\dots,z_{nkD}^{(t)})'$.

The majorizing function given in (\ref{eq:func_h}) is quadratic in each
of $\bs{\mu}$, $\mb{F}$, and $\mb{A}$ when the other two are fixed, and
thus alternating minimization of (\ref{eq:func_h}) with respect to
$\bs{\mu}$ and $\mb{A}$ has closed-form solutions. We now drop the
subscript $(t)$ for notational convenience. For fixed $\mb{F}$ and
$\mb{A}$, set $\bar{z}_{kd}=N_{k}^{-1}\sum_{n=1}^{N}u_{nk}^{*}z_{nkd}$
where $N_{k}=\sum_{n=1}^{N}u_{nk}^{*}$, and write
$\bar{\bs{z}}_{k}=(\bar{z}_{k1},\dots,\bar{z}_{kD})'$. Then the optimal
$\hat{\bs{\mu}}$ is given by
 \begin{align}
	\hat{\bs{\mu}}&=\argmin_{\bs{\mu}}\sum_{n=1}^{N}\sum_{k=1}^{K}u_{nk}^{*}
	 \|\bs{z}_{nk}-(\bs{\mu}+\mb{A}\bs{f}_{k})\|^{2}\notag\\
	 &=N^{-1}\sum_{k=1}^{K}N_{k}(\bar{\bs{z}}_{k}-\mb{A}\bs{f}_{k}).
	 \label{eq:update_mu}
 \end{align}

Optimization of $\mb{F}$ requires a numerical procedure because of its
orthonormality. To update $\mb{F}$ for fixed $\bs{\mu}$ and $\mb{A}$, we
apply the gradient projection (GP) algorithm with the orthonormal
constraint (Jennrich, 2001, 2002). The only problem specific thing
required for the GP algorithm is the gradient of (\ref{eq:func_h})
viewed as a function of $\mb{F}$. Let
$\bar{z}_{kd}^{*}=N_{k}^{-1}\sum_{n=1}^{N}u_{nk}(z_{nkd}-\mu_{d})$, and
write $\bar{\mb{Z}}^{*}=(\bar{z}_{kd}^{*})$. Furthermore, let $\mb{N}$
be a $K\times K$ diagonal matrix where the $k$th diagonal element is
$N_{k}$. Then, the gradient of $h$ at $\mb{F}$ is given as follows:
 \begin{equation}
	\mb{\Gamma}=\frac{\partial
	 h}{\partial\mb{F}}=\frac{1}{4}\mb{N}(\mb{F}\mb{A}'-\bar{\mb{Z}}^{*})\mb{A}.
	 \label{eq:gradient}
 \end{equation}
Using $\mb{\Gamma}$ as the gradient in the GP algorithm with orthonormal
constraint, we obtain the optimal $\hat{\mb{F}}$.

Finally, for fixed $\bs{\mu}$ and $\mb{F}$, the $dl$th element $a_{dl}$
of $\mb{A}$ is updated by solving the minimization problem in
(\ref{eq:func_h}) directly.  Let
$v_{dl}=\sum_{n=1}^{N}\sum_{k=1}^{K}u_{nk}^{*}(z_{nkd}-\mu_{d})f_{kl}$
and $w_{ll'}=\sum_{k}^{K}N_{k}f_{kl}f_{kl'}$. Then, up to a constant,
the loss function with respect to $\mb{A}$ can be written as
 \begin{align}
	h'(\mb{A})
	&=\frac{1}{8}\sum_{d=1}^{D}\sum_{l=1}^{L}\sum_{l'=1}^{L}w_{ll'}a_{dl}a_{dl'}
	 -\frac{1}{4}\sum_{d=1}^{D}\sum_{l=1}^{L}v_{dl}a_{dl}
	+N\sum_{l=1}^{L}\lambda_{l}\sum_{d=1}^{D}|a_{dl}|.
 \end{align}
 Let $s_{dl}=\sign(a_{dl})$ for $a_{dl}\neq0$, and $s_{dl}\in[-1,1]$ for
 $a_{dl}=0$. Thus, the subdifferential $\partial h'_{dl}(\mb{A})$ of
 $h'(\mb{A})$ at $a_{dl}$ is as follows:
 \begin{equation}
	\partial h'_{dl}(\mb{A})=
	 \left\{\frac{1}{4}\sum_{l'=1}^{L}w_{ll'}a_{dl'}-\frac{1}{4}v_{dl}+N\lambda_{l}s_{dl}\right\}.
 \end{equation}
 Then, the optimal $\hat{a}_{dl}$ can be obtained by
 \begin{equation}
	\hat{a}_{dl}=\frac{1}{w_{ll}}\sign(c_{dl})\max(0,
	 |c_{dl}|-4N\lambda_{l}),
	 \label{eq:update_A}
 \end{equation}
 where $c_{dl}=-\sum_{l'\neq l}a_{dl'}+v_{dl}$.

 The procedure of the proposed optimization algorithm is summarized as
 follows:
\begin{enumerate}
 \setlength{\itemindent}{15pt}
 \item[{\it STEP1}.]  Set $t=1$ and initial values of $\bs{\xi}^{(1)}$,
								 $\bs{\mu}^{(1)}$, $\mb{F}^{(1)}$, and $\mb{A}^{(1)}$.
 \item[{\it STEP2}.] Calculate the conditional expectation of $u_{nk}$
							using (\ref{eq:cond_expect}).
 \item[{\it STEP3}.] Update $\bs{\xi}$ using (\ref{eq:update_xi}) and
							set $\bs{\xi}^{(t+1)}=\hat{\bs{\xi}}$.
 \item[{\it STEP4}.] Update $\bs{\mu}$ using (\ref{eq:update_mu}) and
							set $\bs{\mu}^{(t+1)}=\hat{\bs{\mu}}$.
 \item[{\it STEP5}.] Update $\mb{F}$ by the GP algorithm with the
							gradient $\mb{\Gamma}$ in (\ref{eq:gradient}) and set
								 $\mb{F}^{(t+1)}=\hat{\mb{F}}$.
 \item[{\it STEP6}.] Update $\mb{A}$ using (\ref{eq:update_A}) and set
								 $\mb{A}^{(t+1)}=\hat{\mb{A}}$.
 \item[{\it STEP7}.] Increase the value of $t$ by 1 and repeat STEP2-6
							until the penalized log-likelihood
							(\ref{eq:penLL}) converges.
\end{enumerate}

Prior to applying the above algorithm, the value of the regularization
parameters, $\bs{\lambda}=(\lambda_{1},\dots,\lambda_{L})'$, should be
determined. In regression analysis, the degree of freedom for the
shrinkage method (Zou et al, 2007; Hirose et al., 2013) may be used for
selecting the model selection criteria. In this paper, we choose
$\bs{\lambda}$ by minimizing the following Bayesian information
criterion (BIC):
\begin{equation}
 \text{BIC} (\lambda)=-2\ell(\bs{\xi},\bs{\mu},\mb{F},\mb{A})
	+ (\log N) \df(\bs{\lambda}),
	\label{eq:BIC}
\end{equation}
where $\df(\bs{\lambda})$ is the number of nonzero parameters for fixed
$K$ and $L$. The degree of freedom $\df(\bs{\lambda})$ used in
Eq. (\ref{eq:BIC}) is defined as
$\df(\bs{\lambda})=K+D+KL+|\mb{A}|_{\bs{\lambda}}$, where $K$ and $D$
are the length of the vector $\bs{\xi}$ and $\bs{\mu}$, respectively,
$KL$ is the total number of elements of $\mb{F}$, and
$|\mb{A}|_{\bs{\lambda}}$ is the number of nonzero elements of $\mb{A}$
when the regularization parameter is $\bs{\lambda}$. In the following
sections, we use the above BIC to choose $\bs{\lambda}$ in the proposed
method. In addition, although different parameters can be used for
different component loading vectors, we consider using only a single
regularization parameter $\lambda$ for all loadings.

\section{Numerical examples}
\label{sec:simulation}

\subsection{A Monte Carlo simulation}

We conducted a simulation study to evaluate the performance of the
proposed method, compared with tandem analysis (TA), in which sparse
logistic principal component analysis (SLPCA) (Lee et al., 2010) is
conducted, followed by the ordinary $k$-means clustering of estimated
principal component scores.

The artificial data $\mb{Y}$ were generated through the CLUSBIRD model
(\ref{eq:prob_y}) with three clusters ($K=3$) and two dimensional
structure ($L=2$). That is, an object $y_{nd}$ that was assigned to
cluster $k$ was generated by $y_{nd}\sim Ber(\pi_{kd})$. To determine
the value of $\pi_{kd}$, the values of $\bs{\mu}$, $\mb{F}$, and
$\mb{A}$ were generated. We used a zero vector for $\bs{\mu}$. Each
centroid $\bs{f}_{k}$ of clusters in the two-dimensional space were
randomly generated so that the distance between two clusters was equal
for all combinations of two clusters, and then the
$\mb{F}=(\bs{f}_{1},\bs{f}_{2},\bs{f}_{3})$ was orthonormalized. The
loading matrix $\mb{A}$ was set at
\begin{equation*}
 \mb{A}=
	\begin{pmatrix}
	 c\cdot\bs{1}_{D_{1}} & \bs{0}_{D_{1}} \\
	 \bs{0}_{D_{1}} & c\cdot\bs{1}_{D_{1}} \\
	 \bs{0}_{D_{2}} & \bs{0}_{D_{2}}
	\end{pmatrix},
\end{equation*}
where $\bs{1}_{m}$ and $\bs{0}_{m}$ denote $m$-vectors of ones and
zeroes, respectively. Here, $c$ is a scalar whose value was determined
based on sample size as described below. In this simulation study, we
considered three factors: sample size ($N=100,\;300$), the number of
variables ($D=10,\;1000$), and the proportion of informative variables
on the cluster structure ($m=0.5,\;1.0$). Then, we set the value of $c$
was set at 2.5 for $D=10$ and 0.5 for $D=1000$. The number $D_{1}$ was
calculated as $D_{1}=\lfloor\frac{m}{2}D\rfloor$, where
$\lfloor\cdot\rfloor$ denotes a floor function. Thus, based on the above
structure of $\mb{A}$, 2$D_{1}$ variables contributes the
low-dimensional structure and $D_{2}(=D-2D_{1})$ variables are random
error variables. For each condition, we generated 50 replications, thus
yielding $2\times2\times2\times50=400$ random samples in total. We used
the Adjusted Rand Index (ARI) (Hubert and Arabie, 1985) to assess the
recovery of cluster memberships. The ARI has a maximal value of 1 in the
case of a perfect recovery of the underlying cluster structure, and a
value of 0 in the case where the true and estimated class assignments
coincide no more than would be expected by chance.

In this study, we used 50 sets of random initial values of all
parameters for the proposed model and SLPCA, except that initial values
of $\bs{\mu}$ in the proposed model and low-dimensional means in SLPCA
were both set at zero. Also, we used the parameter values of $K$ and $L$
as their values, i.e., values of 3 and 2, respectively, for the two
models. The values of tuning parameter $\lambda$ in the SLPCA model were
determined by BIC as defined in Lee et al. (2010). To reduce
computational burden, we selected the values of tuning parameters only
in the first replication for each condition, and then used the values of
the parameter obtained from the selection by BIC for the remaining
replications.

\begin{figure*}[!tb]
 \begin{center}
\renewcommand{\arraystretch}{1}
 \vspace{0.5cm}
	\begin{tabular}{cc}
	 $D=10$\\
	 \includegraphics[width=15cm]{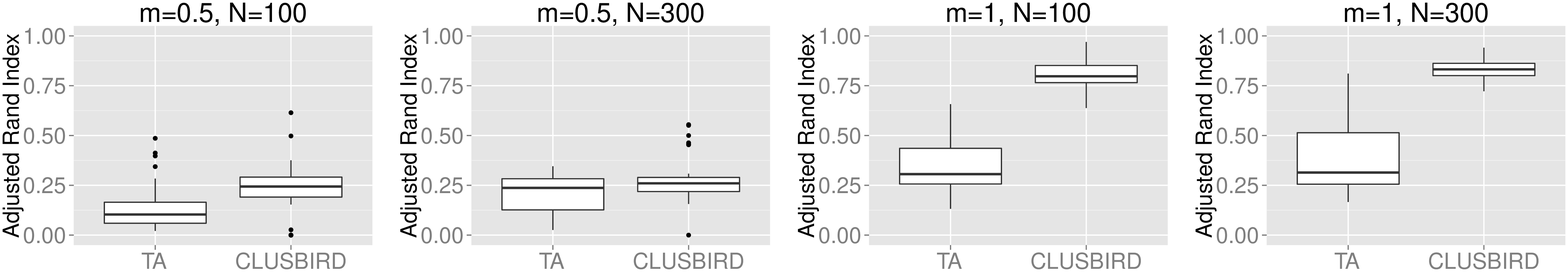}\\
	 $D=1000$\\
	 \includegraphics[width=15cm]{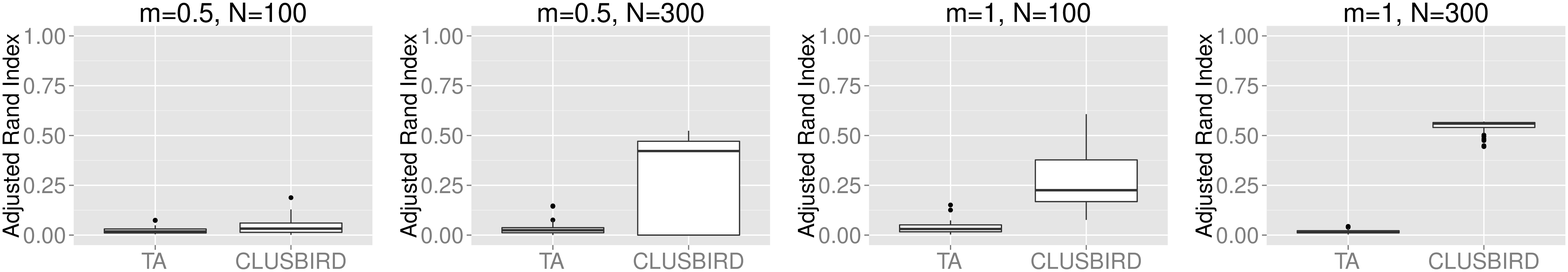}
	 \end{tabular}
 \caption{Boxplots of adjusted Rand indices}
 \label{fig:simulation_boxplot}
\end{center}
\end{figure*}
Figure \ref{fig:simulation_boxplot} shows boxplots of the ARIs obtained
from the two methods, along with the values of $D$, $m$, and $N$. Each
boxplot denotes the values of ARIs for 50 replications under each
condition. When the number of variables is small ($D=10$), the proposed
method provided better results than tandem analysis under all cases. We
can see that the recovery of the cluster structure became better when
the sample size and/or the proportion of the informative variables
increased. Also, under the moderately high-dimensional settings
($D=1000$), the recoveries of the proposed method were superior or
similar to those of tandem analysis. Specifically, tandem analysis did
not work well under the conditions with $D=1000$.

\subsection{Binary image classifications}
\label{sec:binary_image}

\begin{figure*}[!tb]
 \begin{center}
\renewcommand{\arraystretch}{1}
 \vspace{0.5cm}
	\begin{tabular}{c}
	 \includegraphics[width=15cm]{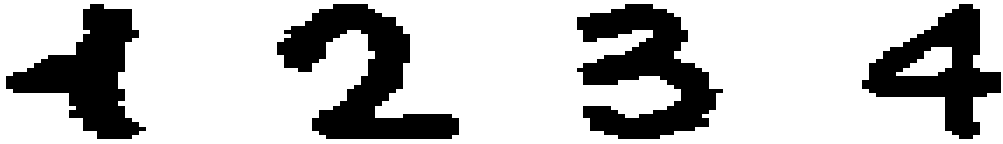}
	 \end{tabular}
 \caption{Examples of normalized bitmaps}
	\label{fig:example_optdigits}
\end{center}
\end{figure*}
Handwritten digit recognition has many application scenarios such as
auto-mail classification according to zip code and signature recognition
(Bouguila, 2010). We used binary image data that were available from the
well-known UCI database (Bache and Lichman, 2013) which contains 5,620
objects. Each object represents one of the integers from 0 to 9, and we
used images of 1, 2, 3, and 4, for which examples are shown in Figure
\ref{fig:example_optdigits}. Each normalized bitmap includes a
$32\times32$ matrix, i.e., a 1,024-dimensional binary vector, in which
each element indicates one pixel with a value of white or black. Fifty
objects for each number were selected and thus $50\times4=200$ objects
were analyzed by the proposed method and tandem analysis with $K=4$ and
$L=2$. For the proposed method and the tandem approach, the value of a
tuning parameter $\lambda$ was determined by BIC.

\begin{figure*}[!tb]
 \begin{center}
\renewcommand{\arraystretch}{1}
 \vspace{0.5cm}
	\begin{tabular}{cc}
	 CLUSBIRD (ARI = 0.72) & Tandem Analysis (ARI = 0.45)\\
	 \includegraphics[width=7cm]{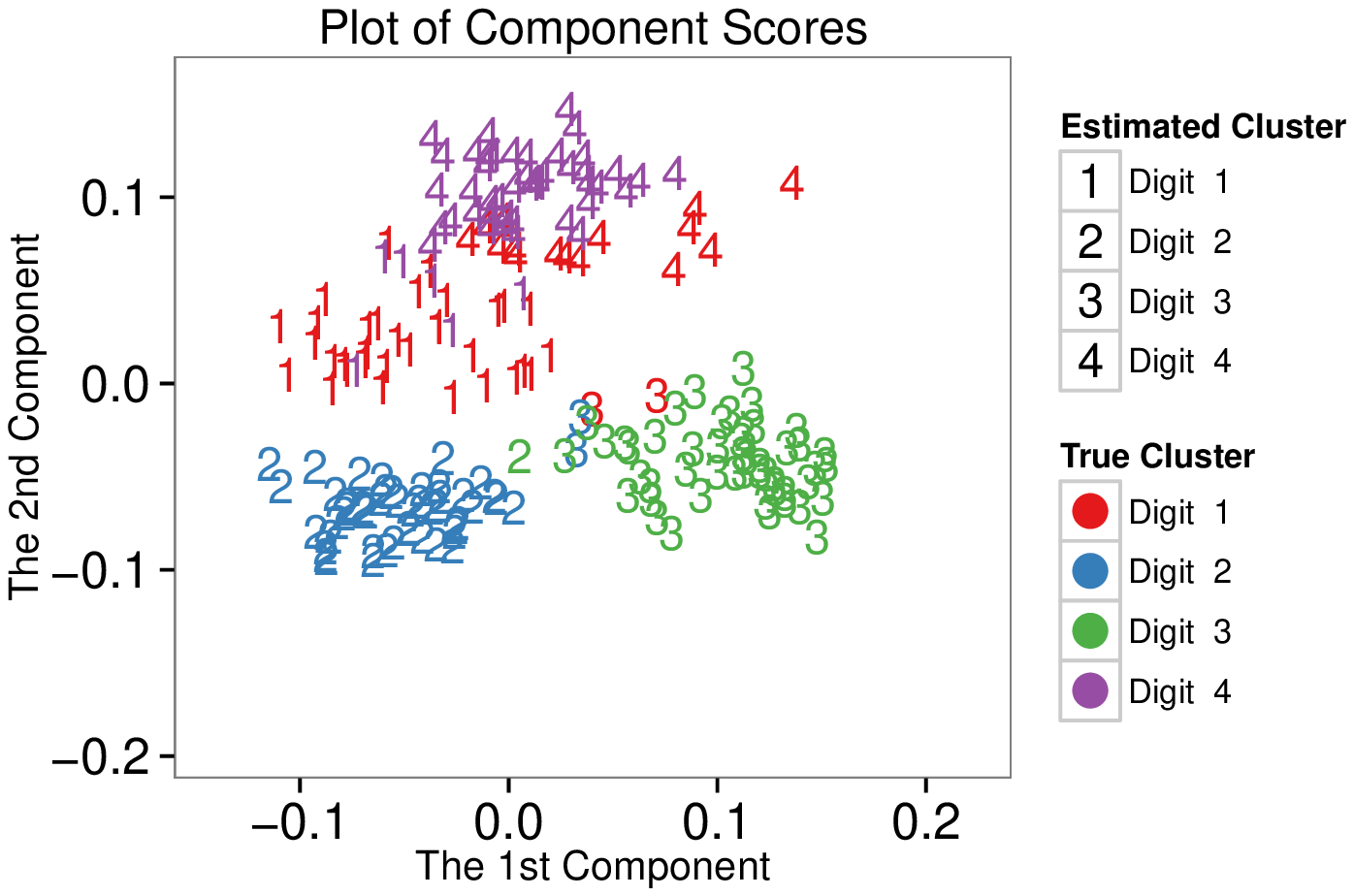} & \includegraphics[width=7cm]{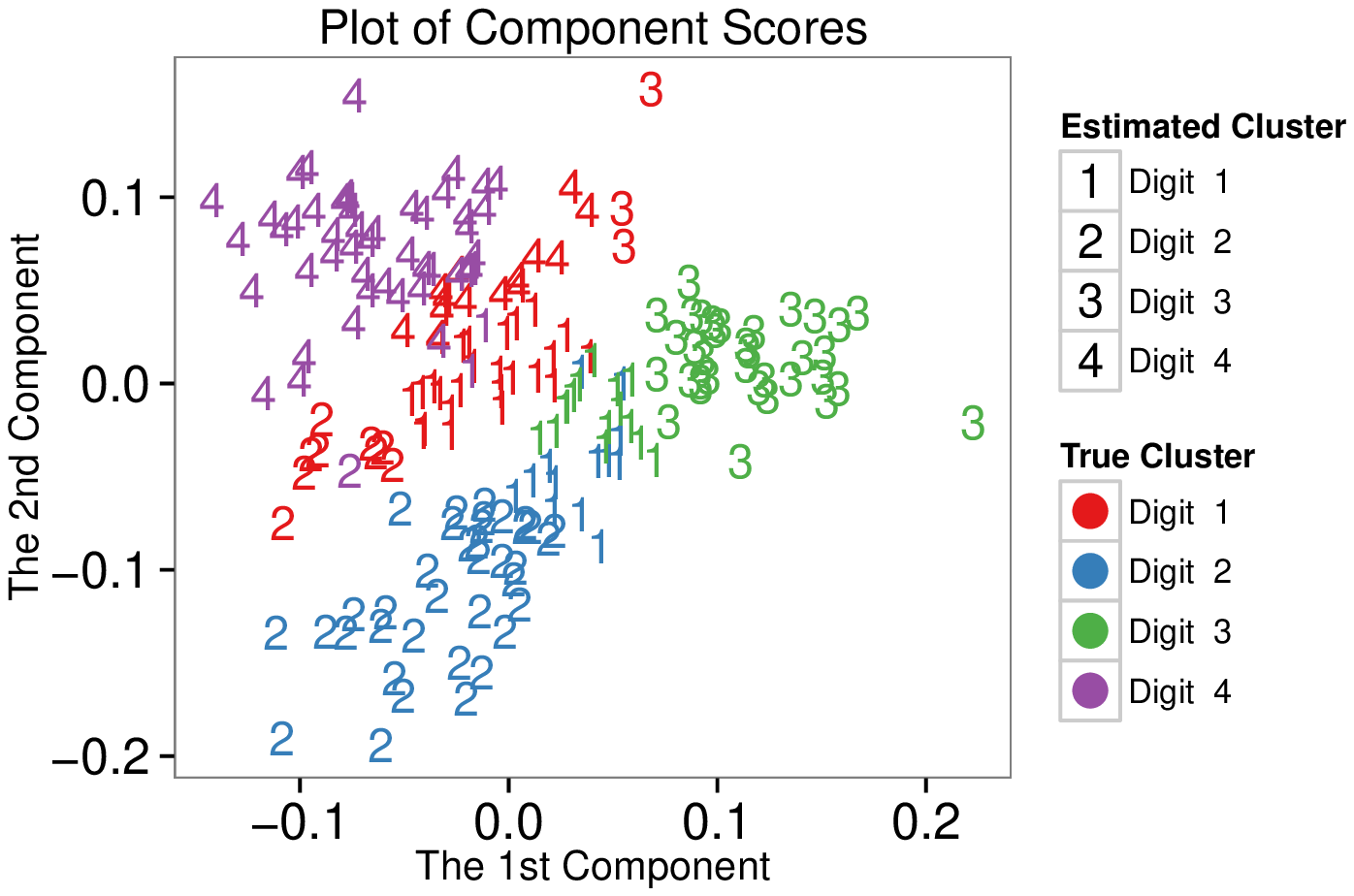}\\
	 \end{tabular}
 \caption{Plots of component scores estimated by CLUSBIRD and tandem
	analysis; in the plots, the number denotes the estimated cluster and
	the color denotes the true cluster} \label{fig:image_plots_F}
\end{center}
\end{figure*}
Estimated component scores with clusters are shown in Figure
\ref{fig:image_plots_F}. It can be seen that the proposed method
provided a well-separated and compact low-dimensional cluster
structure. On the other hand, tandem analysis provided crude recovery of
the true cluster structure. In fact, the value of ARI for the proposed
method was higher than that for tandem analysis. From this viewpoint,
CLUSBIRD provided a better result than that of tandem analysis.

\subsection{Population classification using single nucleotide polymorphism data}
\label{sec:SNPs}

Association studies based on high-throughput single nucleotide
polymorphism (SNP) data have become a popular way to detect genomic
regions associated with complex human diseases. A crucial issue in
association studies is population stratification detection (Hao et al.,
2004), which is to determine whether a population is homogeneous or has
hidden structures within it. With the presence of population
stratification, a naive case-control approach that did not consider the
stratification would yield biased results and, therefore, draw
inaccurate scientific conclusions (Ewens and Spielman, 1995). We used
the SNP dataset available in the International HapMap project (The
International HapMap Consortium, 2005), filtering out those with minor
allele frequencies greater than 0.01 and those missing genotype rates
less than 0.05. The dataset consists of 3 different ethnic populations
of 90 Asians (45 Han Chinese in Beijing, China; CHB and 45 Japanese in
Tokyo, Japan; JPT), 60 Caucasians (Utah residents with ancestry from
northern and western Europe; CEO), and 60 Africans (Yoruba in Ibadan,
Nigeria; YRI). Here, we conducted the proposed method and tandem
analysis to detect the three-subpopulation structure using the SNP data
on the 210 subjects.

Since there were too many SNPs (2.2 million, 2.3 million, and 2.6
million SNPs for CHB-JPT, CEO, and YRI populations, respectively) to
analyze those data, we had to select SNPs that were seen to be
associated with detection of the subpopulation. First, using PLINK
(Purcell, 2007), we conducted three association analyses in which each
population was considered as a case and the other two populations were
control. Then, we obtained SNPs which had genome-controlled p-values
less than $0.1\%$. All those SNPs were considered to be related to the
differences among the three ethnic populations. After selecting SNPs with no
missing values, we finally obtained 589 SNPs of 210 subjects.

\begin{figure*}[!tb]
 \begin{center}
\renewcommand{\arraystretch}{1}
 \vspace{0.5cm}
	\begin{tabular}{c}
	 CLUSBIRD (ARI = 1.00)\\
	 \includegraphics[width=15cm]{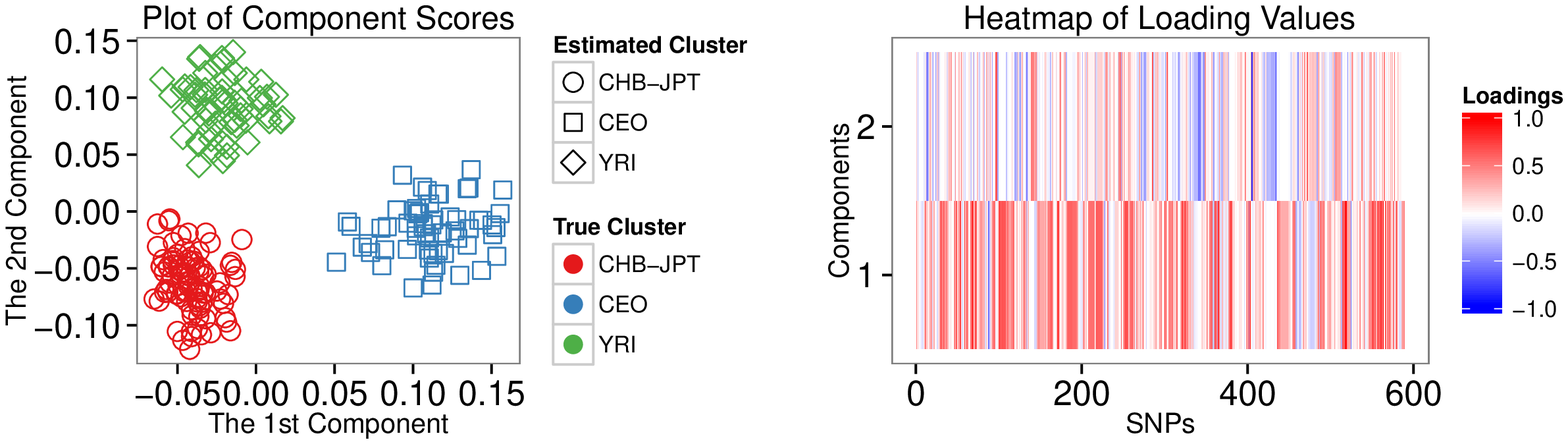}\\
	 Tandem Analysis (ARI = 0.47)\\
	 \includegraphics[width=15cm]{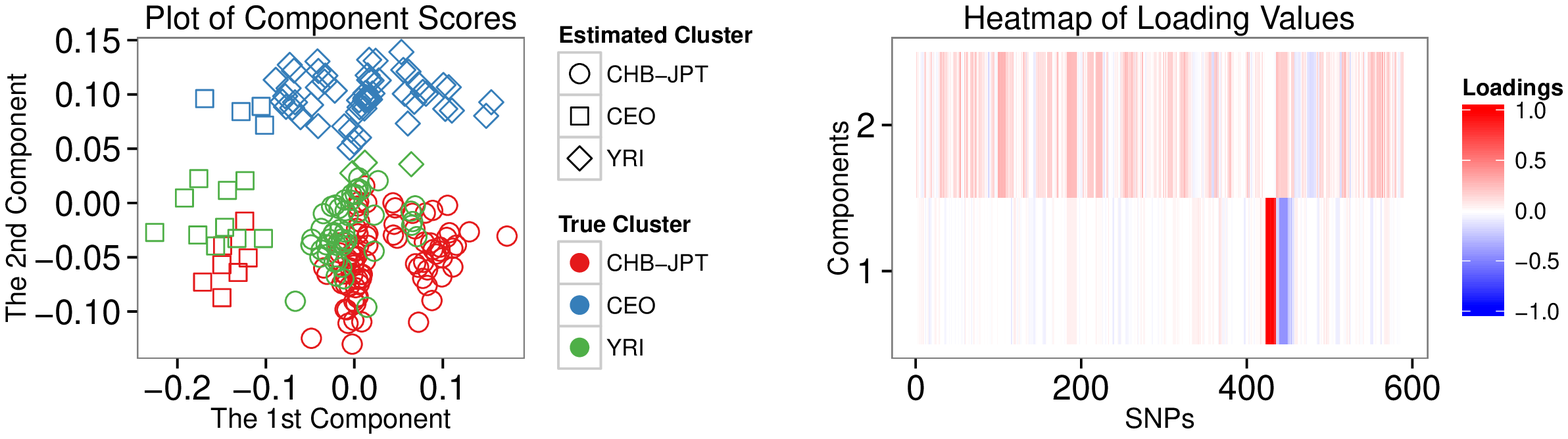}
	 \end{tabular}
 \caption{Plots of component scores (left) and loading values (right)
	estimated by CLUSBIRD and tandem analysis; in the left panel, the
	colors and shapes denote the true and estimated memberships,
	respectively; loading values were scaled so that the value existed in
	$[-1,1]$} \label{fig:snp_plots_F_A}
\end{center}
\end{figure*}
We conducted the proposed CLUSBIRD method and tandem analysis with $K=3$
and $L=2$ using the SNP data. A tuning parameter was determined by BIC
for both methods. The results are shown in Figure
\ref{fig:snp_plots_F_A}. We can see that the proposed method recovered
the true ethnic populations perfectly. In contrast, tandem analysis
provided a crude recovery of the populations. The tandem analysis,
SLPCA, provided a bit sparse estimation of loading values, where only a
few SNPs had large loading values for the first component and many SNPs
had low loading values for the second component. Although this sparse
structure may provide easy interpretation for the estimated
low-dimensional structure, the structure did not contain the true ethnic
populations well. The proposed method provided a reasonably sparse
structure of loading values. Actually, all SNPs used for this analysis
had some relation to detection of populations. Thus, it is reasonable
that all SNPs had large loading values. In addition, for the proposed
method, almost all SNPs had high loading values for one component,
resulting in easy interpretation of the low-dimensional structure.

\section{Conclusion}
\label{sec:conclusion}

In this paper, we proposed a new procedure, called CLUSBIRD, for
simultaneously finding the optimal cluster structure for multivariate
binary objects and finding the subspace to represent the cluster
structure. The proposed method can provide the weight for each binary
variable, which indicates the contribution of the variable to the
cluster structure. In general, tandem analysis for clustering objects
with dimension reduction is likely to fail in finding the cluster
structure. In fact, our numerical examples demonstrate the inability of
tandem analysis to detect the cluster structure and subspace for the
structure. Those examples also show that our proposed method can provide
a better cluster structure than tandem analysis. Furthermore, from the
examples, we found that our procedure can work well for data that had a
mildly larger number of variables than the sample size.

The proposed model can be considered an extension of the ordinary latent
class analysis (LCA) (Aitkin et al., 1981). However, the ordinary LCA
cannot provide loading values for variables and a low-dimensional
structure. Also, LCA may not provide an appropriate estimation with the
moderately high-dimensional dataset we used in the numerical
examples. From this point of view, the proposed method can provide
useful insight for researchers.

The proposed method can be extended to deal with various problems. For
example, it is useful for the proposed model to deal with categorical
variables, not just binary variables. In addition, the ordinary LCA
model is ready for multi-group analysis, the analysis with covariates,
and analysis of repeated measures data (Collins and Lanza, 2009). Using
the formulation of LCA, the proposed model can also contain those
features. These could be interesting topics for further research.

\appendix
\renewcommand{\theequation}{A.\arabic{equation}}
\setcounter{equation}{0}
 \section*{Appendix A: Estimation of individual latent scores}
\label{sec:estimate_scores}

To obtain individual component scores, $\mb{G}=(g_{nl})$, we propose a
two-step approach. First, we estimate all parameters, $\bs{\mu}$,
$\mb{A}$, $\mb{F}$, and $\bs{\xi}$, in the CLUSBIRD model. Then, we
assume that a cluster structure of individuals is present in a
low-dimensional space that is the same as that for the cluster center
$\mb{F}$. That is, the estimated loading matrix $\hat{\mb{A}}$ and
low-dimensional centroids $\hat{\bs{\mu}}$ also define the subspace for
the individuals. Thus, we consider the following post hoc model. Suppose
that $\tilde{y}_{nd}$ ($n=1,\dots,N;\;d=1,\dots,D$) follows the
Bernoulli distribution with success probability
$\pi_{nd}=\pi(\theta_{nd})$, where $\theta_{nd}$ is the logit
transformation of $\pi_{nd}$. In addition, we assume that the canonical
parameter $\theta_{nd}$ has a low-rank representation
\begin{equation*}
 \theta_{nd}=\hat{\mu}_{d} + \bs{g}_{n}'\hat{\bs{a}}_{d},
\end{equation*}
where $\bs{g}_{n}=(g_{n1},\dots,g_{nL})'$ with
$\mb{G}'\mb{G}=\mb{I}_{L}$. Here, we write
\begin{equation*}
 S(\mb{G})=\sum_{n=1}^{N}\sum_{d=1}^{D}\log\pi(q_{nd}(\hat{\mu}_{d}+
	\bs{g}_{n}'\hat{\bs{a}}_{d})).
\end{equation*}
Then, we obtain individual component scores by maximizing $S(\mb{G})$
over $\mb{G}$. Similar to the solution of $\mb{F}$ in Section 3, the
optimal $\mb{G}$ can be obtained using the GP algorithm.


\begin{thebibliography}{99}
 \bibitem[Aitkin (1981) Aitkin]{AitkinAnderson1981} Aitkin, M.,
								 Anderson, D., and Hinde, J. (1981). Statistical modeling
								 of data on teaching styles. \textit{Journal of the
								 Royal Statistical Society, Series A}, 144, 419--461.
 \bibitem[Arabie (1994) Arabie]{ArabieHubert1994} Arabie, P. and Hubert,
								 L. (1994). Cluster analysis in marketting research. In
								 Bagozzi, R.P., editor, \textit{Advanced methods of
								 marketing research} (pp.160--189). Blackwell, Oxford.
 \bibitem[Bache (2013)]{BacheLichman2013} Bache, K. and Lichman,
								 M. (2013). UCI Machine Learning
					Repository [http://archive.ics.uci.edu/ml]. Irvine, CA:
					University of California, School of Information and Computer
								 Science. Accessed Apr. 24, 2014.
 \bibitem[Bouguila (2010) Bouguila]{Bouguila2010} Bouguila,
								 N. (2010). On multivariate binary data clustering and
								 feature weighting. \textit{Computational Statistics \&
								 Data Analysis}, 54, 120--134.
 \bibitem[Browne (2001)]{Browne2001} Browne, M.W. (2001). An overview of
								 analytic rotation in exploratory factor
								 analysis. \textit{Multivariate Behavioral Research},
								 36, 111--150.
 \bibitem[Cagnone (2012) Cagnone]{CagnoneViroli2012} Cagnone, S. and
								 Viroli, C. (2012). A factor mixture analysis model for
								 multivariate binary data. \textit{Statistical
								 Modelling}, 12, 257--277.
 \bibitem[Collins (2002) Collins]{CollinsDasgupta2002} Collins, M.,
								 Dasgupta, S., and Schapire, R.E. (2002). A
								 generalization of principal component analysis to the
								 exponential family. In \textit{Advanced in Neural
								 Information Processing System} (T.G. Dietterich,
								 S. Becker, and Z. Ghahramani, eds.), 14, 617--642. MIT
								 Press, Cambridge, MA.
 \bibitem[Collins (2010) Collins]{Collins2010} Collins, L.M. and Lanza,
								 S.T. (2010). Latent class and latent transition
								 analysis with applications in the social, behavioral,
								 and health sciences. John Wiley \& Sons, Inc., New
								 Jersey.
 \bibitem[deLeeuw (2006) deLeeuw]{deLeeuw2006} De Leeuw, J. (2006).
								 Principal component analysis of binary data by iterated
								 singular value decomposition. \textit{Computational
								 Statistics \& Data Analysis}, 50, 21--39.
 \bibitem[De Soete et al. (1994) De Soete and Carroll]{DeSoete1994} De
								 Soete, G. and Carroll, J.D. (1994). K-means clustering
								 in a low-dimensional Euclidean space. In Diday, E. and
								 Lechevallier, Y. and Schader, M. and Bertrand, P. and
								 Burtschy, B. (Eds.) New Approaches in Classification
								 and Data Analysis (pp. 212-219). Springer, Heidelberg
 \bibitem[Dempster (1977) Dempster]{DempsterLaird1977} Dempster, N.M.,
								 Laird, A.P., and Rubin, D.B. (1977). Maximum likelihood
								 from incomplete data via the EM algorithm (with
								 discussion), \textit{Journal of the Royal Statistical
								 Society B}, 39, 1--38.
 \bibitem[DeSarbo (1990) DeSarbo]{DeSarbo1990} DeSarbo, W.S., Jedidi,
								 K., Cool, K., and Schendel, D. (1990). Simultaneous
								 multidimensional unfolding and cluster analysis: An
								 investigation of strategic groups. \textit{Marketing
								 Letters}, 2, 129--146.
 \bibitem[Ekholm (2000)]{EkholmMcDonald2000} Ekholm, A., McDonald, J.W.,
								 and Smith, P.W.F. (2000). Association models for a
								 multivariate binary response. \textit{Biometrics}, 56,
								 712--718.
 \bibitem[Ewens (1995)]{EwensSpielman1995} Ewens, W.J. and Spielman,
								 R.S. (1995). The transmission/disequilibrium test:
								 History, subdivision, and admixture. \textit{The
								 American Journal of Human Genetics}, 57, 455--464.
 \bibitem[Ghahramani (1997) Ghahramani]{Ghahramani1997} Ghahramani,
								 Z. and Hilton, G.E. (1997). The EM algorithm for
								 mixture of factor analyzers. Technical Report
								 CRG-TR-96-1, Department of Computer Science, University
								 of Toronto, Canada.
 \bibitem[Hao (2004)]{HaoLi2004} Hao, K., Li, C., Rosenow, C., and Wong,
								 W.H. (2004). Detect and adjust for population
								 stratification in population-based association study
								 using genomic control markers: An application of
								 Affymetrix Genechip${}^{\textregistered}$ Human Mapping
								 10K array. \textit{European Journal of Human Genetics},
								 12, 1001--1006.
 \bibitem[Hirose (2013)]{HiroseTateishi2013} Hirose, K., Tateishi, S.,
								 and Konishi, S. (2013). Tuning parameter selection in
								 sparse regression modeling. \textit{Computational
								 Statistics \& Data Analysis}, 59, 28--40.
 \bibitem[Hirose (2014)]{HiroseYamamoto2014} Hirose, K., and Yamamoto,
								 M. (2014). Sparse estimation via nonconcave penalized
								 likelihood in a factor analysis
								 model. \textit{Statistics and Computing}, in press.
 \bibitem[Hunter (2004) Hunter]{HenterLange2004} Hunter, D.R. and Lange,
								 K. (2004). A tutorial on MM algorithms. \textit{The
								 American Statistician}, 58, 30--37.
 \bibitem[Jaakkola (2000) Jaakkola]{JaakkolaJordan2000} Jaakkola,
								 T.S. and Jordan, M.I. (2000). Bayesian parameter
								 estimation via variational methods. \textit{Statistics
								 and Computing}, 10, 25--37.
 \bibitem[Jennrich (2001) Jennrich]{Jennrich2001} Jennrich, R.I. (2001).
								 A simple general procedure for orthogonal
								 rotation. \textit{Psychometrika}, 66, 289--306.
 \bibitem[Jennrich (2002) Jennrich]{Jennrich2002} Jennrich, R.I. (2002).
								 A simple general procedure for oblique
								 rotation. \textit{Psychometrika}, 67, 7--20.
 \bibitem[Juan (2004) Vidal]{JuanVidal2004} Juan, A. and Vidal,
								 E. (2004). Bernoulli mixture models for binary
								 images. Proceedings of the ICPR 2004.
 \bibitem[Kozma (2009)]{KozmaIlin2009} Kozma, L., Ilin, A., and Raiko,
 								 T. (2009). Binary principal component analysis in the
 								 Netflix collaborative filtering
 								 task. \textit{Proceedings of 2009 IEEE International
 								 Workshop on Machine Learning for Signal Processing}.
 \bibitem[Lee (2010) Lee]{LeeHuang2010} Lee, S. and Huang, J.Z. and Hu,
								 J. (2010). Sparse logistic principal components analysis
								 for binary data. \textit{The Annals of Applied
								 Statistics}, 4, 1579--1601.
 \bibitem[Milligan (1996)]{Milligan1996} Milligan,
								 G.W. (1996). Clustering validation: Results and
								 implications for applied analysis. In: Arabie, P.,
								 Hubert, L.J., De Soete, G. (Eds.), Clustering and
								 Classification. World Scientific Publishing, River
								 Edge, pp. 341--375.
 \bibitem[Patrilainen (2004) Patrikainen]{PatrikainenMannila2004}
								 Patrikainen, A. and Mannila, H. (2004). Subspace
								 clustering of high-dimensional binary data - a
								 probabilistic approach. In Workshop on Clustering High
								 Dimensional Data and its Applications, SIAM
								 International Conference on Data Mining. 57--65.
 \bibitem[Purcell (2007)]{PurcellNeale2007} Purcell, S., Neale, B.,
								 Todd-Brown, K., Thomas, L., Ferreira, M.A.R., Bender,
								 D., Maller, J., Sklar, P., de Bakker, P.I.W., Daly,
								 M.J., and Sham, P.C. (2007). PLINK: a toolset for
								 whole-genome association and population-based linkage
								 analysis. \textit{American Journal of Human Genetics}, 81.
 \bibitem[Schein (2003) Schein]{ScheinSaul2003} Schein, A.I., Saul,
								 L.K., and Ungar, L.H. (2003). A generalized linear model
								 for principal component analysis of binary data. In
								 \textit{Proceedings of the Ninth International Workshop
								 on Artificial Intelligence and Statistics} (C.M. Bishop
								 and B.J. Frey, eds.), 38, 14--21. Key West, FL.
 \bibitem[Tamhane (2009) Tamhane]{TamhaneQiu2009} Tamhane, A.C., Qiu,
								 D., and Ankenman, B.E. (2010). A parametric mixture
								 model for clustering multivariate binary
								 data. \textit{Statistical Analysis and Data Mining}, 3,
								 3--19.
 \bibitem[Hapmap (2005)]{Hapmap2005} The International HapMap
								 Consortium. (2005). A haplotype map of the human
								 genome. \textit{Nature}, 437, 1299--1320.
 \bibitem[Tibshirani (1996) Tibshirani]{Tibshirani1996} Tibshirani,
								 R.J. (1996). Regression shrinkage and selection via the
								 lasso. \textit{Journal of the Royal Statistical
								 Society, Series B}, 58, 267--288.
 \bibitem[Timmerman (2010) Timmerman]{TimmermanCeulemans2010} Timmerman,
								 M.E., Ceulemans, E., Kiers, H.A.L., and Vichi,
								 M. (2010). Factorial and reduced k-means
								 reconsidered. \textit{Computational Statistics \& Data
								 Analysis}, 54, 1858--1871.
 \bibitem[Vichi et al. (2001) Vichi and Kiers]{Vichi2001} Vichi, M. and
								 Kiers, H.A.L. (2001). Factorial k-means analysis for
								 two-way data. \textit{Computational Statistics \& Data
								 Analysis}, 37, 49--64.
 \bibitem[Wu (2013)]{Wu2013} Wu, B. (2013). Sparse cluster analysis of
								 large-scale discrete variables with application to
								 single nucleotide polymorphism data. \textit{Journal of
								 Applied Statistics}, 40, 358--367.
 \bibitem[Yamamoto and Hwang (2014)]{YamamotoHwang2014} Yamamoto, M. and
								 Hwang, H. (2014). A general formulation of cluster
								 analysis with dimension reduction and subspace
								 separation. \textit{Behaviormetrika}, 41, 115--129.
 \bibitem[Yoshida (2004) Yoshida]{YoshidaHiguchi2004} Yoshida, R.,
								 Higuchi, T., and Imoto, S. (2004). A mixed factors model
								 for dimension reduction and extraction of a group
								 structure in gene expression data. \textit{Proceedings
								 of the 2004 IEEE Computational Systems Bioinformatics
								 Conference}, 161--172.
 \bibitem[Zou (2007)]{ZouHastie2007} Zou, H., Hastie, T., and
								 Tibshirani, R. (2007). On the degrees of freedom of the
								 lasso. \textit{The Annals of Statistics}, 35,
								 2173--2192.
\end{thebibliography}
\end{document}